\documentclass[preprint,journal]{vgtc}            % preprint (journal style)

\onlineid{9022}

\vgtccategory{Research}

\vgtcpapertype{application/design study}

\title{Towards Explainable Quantum AI: Informing the Encoder Selection of Quantum Neural Networks via Visualization}

\author{%
  \authororcid{Shaolun Ruan}{0000-0002-6163-9786},
  \authororcid{Feng Liang}{0000-0003-3812-3754},
  \authororcid{Rohan Ramakrishna}{0009-0004-6074-4682},
  \authororcid{Chao Ren}{0000-0001-9096-8792},\\
  \authororcid{Rudai Yan}{0000-0001-6777-8368},
  \authororcid{Qiang Guan}{0000-0002-3804-8945},
  \authororcid{Jiannan Li}{0000-0001-8409-4910}
  and \authororcid{Yong Wang}{0000-0002-0092-0793}
}

\authorfooter{
  \item
  	Shaolun Ruan and Jiannan Li are with Singapore Management University.
  	E-mail: slruan.2021@phdcs.smu.edu.sg, jiannanli@smu.edu.sg
  
  \item
  	Feng Liang, Rohan Ramakrishna, Rudai Yan and Yong Wang are with Nanyang Technological University.
  	E-mail: \{feng011\,$|$\,roha0012\,$|$\,rudai011\}@e.ntu.edu.sg, yong-wang@ntu.edu.sg.
    Yong Wang is the corresponding author.

  \item
    Chao Ren is with KTH Royal Institute of Technology.
    E-mail: renc0003@e.ntu.edu.sg

  \item Qiang Guan is with Kent State University.
  	E-mail: qguan@kent.edu.
}

\abstract{%
Quantum Neural Networks (QNNs) represent a promising fusion of quantum computing and neural network architectures, offering speed-ups and efficient processing of high-dimensional, entangled data. A crucial component of QNNs is the encoder, which maps classical input data into quantum states. However, choosing suitable encoders remains a significant challenge, largely due to the lack of systematic guidance and the trial-and-error nature of current approaches.
This process is further impeded by two key challenges: (1) the difficulty in evaluating encoded quantum states prior to training, and (2) the lack of intuitive methods for analyzing an encoder’s ability to effectively distinguish data features. To address these issues, we introduce a novel visualization tool, \emph{\toolName}, which enables QNN developers to compare classical data features with their corresponding encoded quantum states and to examine the mixed quantum states across different classes. By bridging classical and quantum perspectives, \emph{\toolName} facilitates a deeper understanding of how encoders influence QNN performance. Evaluations across diverse datasets and encoder designs demonstrate \emph{\toolName}'s potential to support the exploration of the relationship between encoder design and QNN effectiveness, offering a holistic and transparent approach to optimizing quantum encoders. Moreover, domain experts used \emph{\toolName} to derive two key practices for quantum encoder selection, grounded in the principles of pattern preservation and feature mapping.
}


\teaser{
  \centering
  \includegraphics[width=\linewidth, alt={the interface of \toolName{}}]{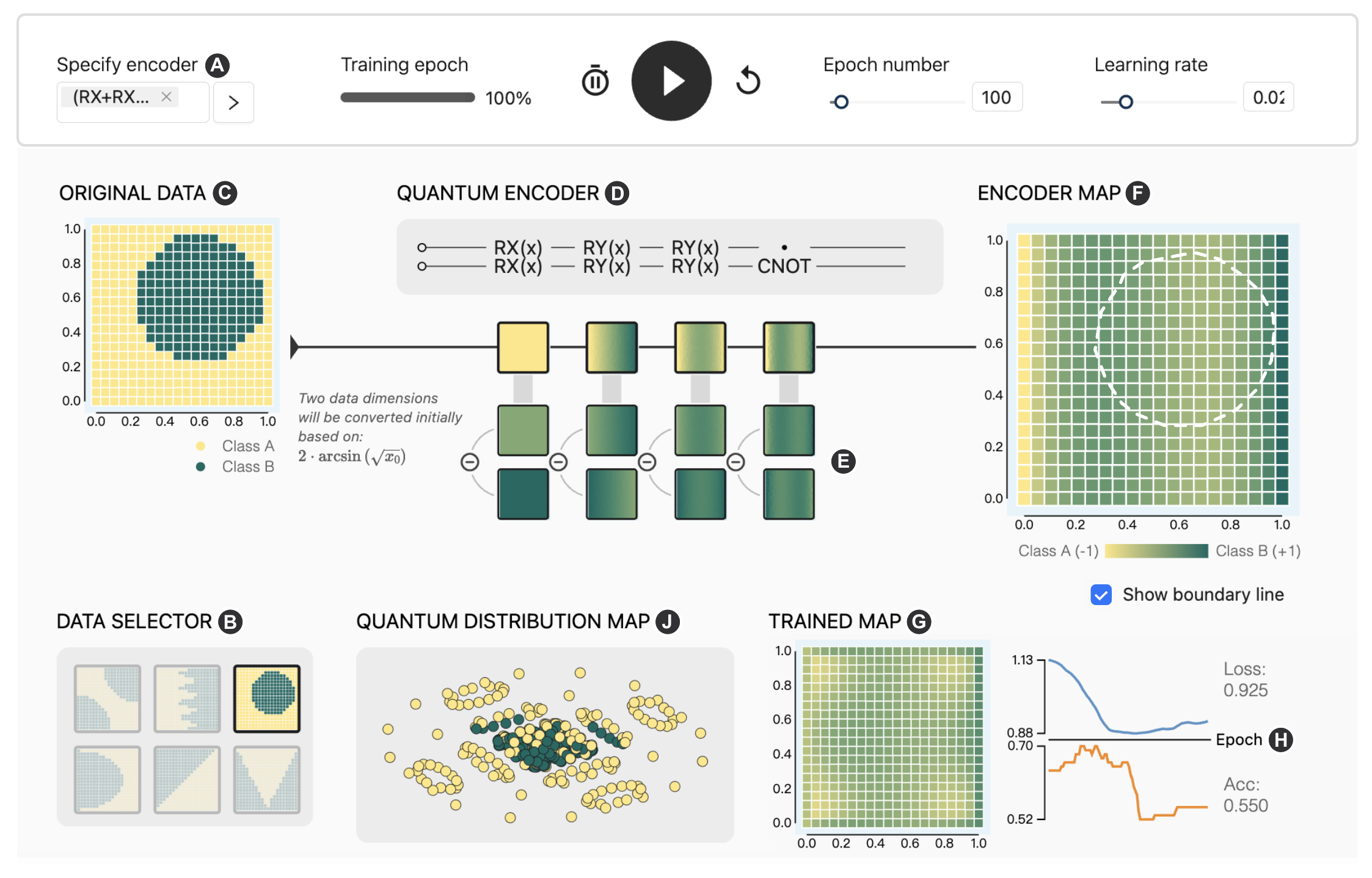}
  \caption{%
The interface of \toolName{} supports the reasoning and selection of the encoder's quality in QNNs. 
The Original Data View (C) visualizes the input dataset. 
The Encoded Data Evolution View (E) illustrates the quantum circuit used for encoding together with the illustration of the encoding process in each step.
The Encoder Map View (F) displays the encoded data as a heatmap, facilitating a direct comparison with the original data. 
The Quantum Distribution Map (J) provides an intuitive representation to show how well the encoder distinguishes the data points with different classes.
The Trained Map View (G) visualizes the final learned patterns after training, while the Performance Analysis View (H) uses line charts to depict the training loss and accuracy. 
  }
  \label{fig:teaser}
}

\graphicspath{{figs/}{figures/}{pictures/}{images/}{./}} % where to search for the images

\usepackage{tabu}                      % only used for the table example
\usepackage{booktabs}                  % only used for the table example
\usepackage{lipsum}                    % used to generate placeholder text
\usepackage{mwe}                       % used to generate placeholder figures

\usepackage{braket}
\usepackage{amsmath}

\usepackage{xcolor}
\usepackage{tikz}

\usepackage{braket}

\usepackage{mfirstuc}

\usepackage{gensymb}
\usepackage{soul}

\newcommand*\component[1]{\tikz[baseline=(char.base)]{
            \node[circle,draw=black, line width=0.5mm,text=white, fill=black, inner sep=0pt,minimum size=11pt](char) {\fontfamily{qhv}\selectfont{#1}}}}

\usepackage{graphicx}
\newcommand{\iconRange}{\raisebox{-.5ex}{\includegraphics[scale=0.25]{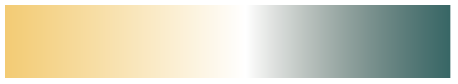}}}%
\newcommand{\iconOptionA}{\raisebox{-.5ex}{\includegraphics[scale=0.2]{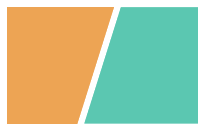}}}%
\newcommand{\iconOptionB}{\raisebox{-.5ex}{\includegraphics[scale=0.2]{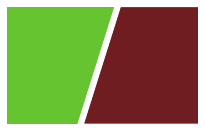}}}%
\newcommand{\iconOptionC}{\raisebox{-.5ex}{\includegraphics[scale=0.2]{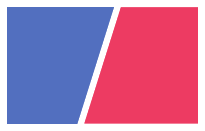}}}%
\newcommand{\iconOptionD}{\raisebox{-.5ex}{\includegraphics[scale=0.2]{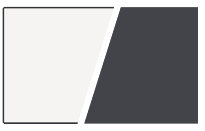}}}%
\newcommand{\iconOptionE}{\raisebox{-.5ex}{\includegraphics[scale=0.2]{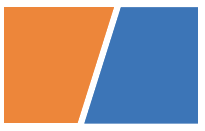}}}%

\newcommand{\revise}[2]{\textcolor{black}{#2}}
\newcommand{\reviseSecond}[1]{\textcolor{black}{#1}}

\newcommand{\toolName}{\textit{XQAI-Eyes}}
\usepackage[multiple]{footmisc}

\usepackage{mathptmx}                  % use matching math font

\begin{document}

%%%%%%%%%%%%%%%%%%%%%%%%%%%%%%%%%%%%%%%%%%%%%%%%%%%%%%%%%%%%%%%%
%%%%%%%%%%%%%%%%%%%%%% START OF THE PAPER %%%%%%%%%%%%%%%%%%%%%%
%%%%%%%%%%%%%%%%%%%%%%%%%%%%%%%%%%%%%%%%%%%%%%%%%%%%%%%%%%%%%%%%

\firstsection{Introduction}

\maketitle

Quantum computing is an emerging field that harnesses the principles of quantum mechanics to achieve advantages beyond the capabilities of classical computation. Building on this foundation, quantum neural networks (QNNs) integrate quantum operations with neural network architectures to tackle complex optimization problems~\cite{liu2021hybrid, nakaji2021quantum, da2017neural}.
\revise{R3-1}{
Compared to their classical counterparts, QNNs demonstrate the potential for exponential speed-ups and are particularly well-suited for processing high-dimensional and entangled data~\cite{yun2023quantum, heidari2022toward, abbas2021power, shen2023quantum}. Furthermore, by exploiting quantum parallelism, QNNs facilitate more efficient learning and optimization processes~\cite{zhang2023quantum, afane2025atp}.
}
Meanwhile, visualization, which has proven to be a suitable scientific educational tool, \reviseSecond{has seen an increasing proliferation in research}~\cite{bethel2023quantum, ruan2023quantumeyes, ruan2024violet, wen2023quantivine} and applications~\cite{quirk, ibmq, qsphere} of quantum computing in recent years, significantly enhancing the transparency of black-box quantum algorithms.

Typically, a QNN consists of three basic components: the encoder, the ansatz, and the measurement. 
As the first component, the encoder plays a crucial role in transforming classical input data into quantum states, enabling the encoded data to be recognized and trained by the subsequent ansatz layer.
Due to the fact that different encoders can produce completely different results, the selection and design of the encoder significantly affect the final performance of the QNN~\cite{mattern2021variational, nguyen2022evaluation}.
More specifically, the model could achieve optimal performance only if the original data features can be effectively transformed into the corresponding quantum states with the preserved features.

\revise{R1-1}{Despite the importance of the QNN encoder, it is still tricky for QNN developers to select the existing encoder templates for their tasks, \reviseSecond{because of the lack of knowledge} to understand how a certain quantum encoder will affect a QNN's performance.}
Consequently, current practice involves repeatedly implementing different encoders until decent performance is achieved.
However, this trial-and-error process has two major problems. 
Firstly, it is highly time-consuming and labor-intensive to find an appropriate encoder. 
Secondly, developers may miss the optimal encoder if they deem some commonly-used encoders satisfactory and settle for them too early.
Therefore, it is particularly necessary for QNN developers to build a cognitive model and enhance their understanding of encoder selection, enabling them to find the most appropriate encoders based on their informed knowledge.
However, there are two significant challenges to address this problem. 
\revise{R1-2}{Firstly, there is rarely work that studied how to \textbf{visualize the encoded classical data points} and further inform the developers the correlation between the original data and post-encoded data.} Secondly, from the angle of explainable AI, there are no methods available to \textbf{intuitively analyze whether the encoder effectively distinguishes the features of different classes}. 
This lack of transparency in the encoder's performance introduces a significant hurdle for developers.

To address the aforementioned challenges, we first proposed a novel quantum circuit measurement method, namely \textit{Encoder Expectation Measurement}, which extracts the expectation value of each original data point and transforms the abstract encoded states into an observable ``classical'' variable.
This method allows users to directly gain knowledge of how different features of data points are encoded within different QNNs.
Additionally, to visually highlight how well the encoder distinguishes between different classes of data, 
we developed a visualization approach called \textit{\revise{R1-Minor}{State Comparison Map}} to intuitively provide a clear overview of the mixed levels of different encoded data,
enhancing the understanding from a ``quantum'' perspective by comparing it with the above ``classical'' method.
With the integration of these two methods, we developed an Explainable Quantum Artificial Intelligence (XQAI) visualization system, namely \toolName{},
allowing users to freely explore and understand how different encoders capture the data features and,
 furthermore, affect the model's performance. 
Users can experiment with various cases (i.e., 6 datasets $\times$ 10 encoders $=$ 60 cases), 
which offers users a holistic view of how to design an optimal encoder regarding the features of training data.
We finally conducted case studies and post-study interviews with quantum computing experts to extensively evaluate \toolName{}. 
To the best of our knowledge, \toolName{} is the first work to visually explain the performance of quantum encoders, bridging the gap between the encoder reasoning and the QNN performance.

The key contributions of this work are summarized as follows:

\begin{itemize}
    \item We first formulated the design requirements for informing QNN developers of the guidance of encoder selection, based on the semi-structured interviews with QNN developers.

    \item We then proposed two methods, i.e., Encoder Expectation Measurement and the Quantum Distribution View, from the perspectives of ``classical'' and ``quantum''.
    These two methods facilitate 1) the comparison between data features and encoded quantum states, and 2) the analysis of the mixed quantum states of different classes, respectively.

    \item Based on the distilled design requirements and the two methods described above, we developed a visualization tool, \toolName{} to enhance QNN developers' understanding and facilitate intuitive insights into how encoders impact QNN model performance.
    
\end{itemize}

\revise{R1-12}{We made \toolName{}\footnote{https://q-encoder-vis.vercel.app/}\textsuperscript{,}\footnote{https://github.com/shaolunruan/XQAI-Eyes}  public to benefit QNN developers.}

\section{Related Work}

Our research relates to
visualization for quantum computing, 
visualization for deep neural network explainability and performance evaluation of quantum data encoders.

\subsection{Visualization for Quantum Computing}

Visualization techniques have been used in quantum computing for quantum circuit visualization and quantum state visualization.

\textit{\reviseSecond{Quantum circuit visualization.}}
Recent work focused on utilizing generally-applicable visualization approaches to enhance the explainability of quantum circuits.  
For example, QuantumEyes~\cite{ruan2023quantumeyes} introduced an interactive visualization system that enhances the interpretability of quantum circuits by mapping network activations and weight distributions onto intuitive graphical representations.
Williams et al.~\cite{williams2021qcvis} developed a sophisticated quantum simulator framework designed to compute qubit state probabilities;
Furthermore, Karafyllidis et al.~\cite{karafyllidis2003visualization} proposed a matrix-like visualization approach to solve the non-transparency problem for quantum-version Fourier Transform algorithm.
Some co-design designs~\cite{kim2025toward} were also proposed to yield the design principles for designing the quantum computing interfaces.

\textit{\reviseSecond{Quantum state visualization.}}
The visualization of quantum states has been explored using various representations of the state vector. 
The  Bloch Sphere representation~\cite{bloch1946nuclear} maps any pure single‐qubit state to a unique point on a three‐dimensional unit sphere.
This fundamental approach has inspired further adaptations; for example, Makela et al.~\cite{makela2010n} extended the Bloch sphere concept to handle two‐qubit states by embedding their joint state space in a higher-dimensional geometric framework.
% while Altepeter et al.~\cite{altepeter2009multiple} proposed methods for visualizing multiple‐qubit systems that preserve intuitive geometric interpretations despite the exponential growth in complexity.
Wille et al.~\cite{wille2021visualizing} introduced a tree-like design that hierarchically decomposes a state vector into constituent components.
In a similar spirit, Zulehner et al.~\cite{zulehner2019efficiently} employed decision diagrams to succinctly represent the matrix form of quantum state amplitudes.
In addition, the VENUS system proposed by Ruan et al.~\cite{ruan2023venus} provides a geometrical representation to explain how these geometric features relate to the underlying quantum behavior.

Despite their effectiveness, none of them could be directly applied to explain the inner workings of quantum neural networks.
The latest work, namely Violet~\cite{ruan2024violet}, was the most relevant study to our work, 
which opened the black box of QNNs by visualizing the quantum state evolution in each component.
However, it did not focus on the encoder part, which was the most critical part of QNNs.
Moreover, we move beyond the transparency enhancement of QNN by unveiling the way how different encoders captured the data features and affected the model's performance accordingly.

\subsection{Visualization for Deep Neural Network Explainability}

Due to the increasing complexity of data and deep neural network models, various visualization approaches were proposed to understand~\cite{ming2018rulematrix, kahng2017Activis, selvaraju2017grad, cheng2021vbridge, wang2020visual}, diagnose~\cite{liu2016towards, wang2019atmseer, tyagi1912navigator, jin2022gnnlens, yang2022diagnosing, sun2020dfseer}, and improve~\cite{schneider2018integrating, xiang2019interactive, yang2022diagnosing} these models.
Specifically, these approaches were categorized into feature-oriented and evolution-oriented visualizations~\cite{liu2018deeptracker}.
Feature-oriented visualizations helped understand the features learned by the model and their impact on predictions. 
For example, Grad-CAM~\cite{selvaraju2017grad} highlighted important areas in input images for convolutional neural network (CNN) predictions. 
VBridge~\cite{cheng2021vbridge} linked influential features with raw data. 
$M^{2}Lens$~\cite{wang2021m2lens} visualized multimodal features, including verbal, acoustic, and visual elements. 
Neuron-level interpretations were also explored by techniques like NeuroCartography~\cite{park2022NeuroCartograph}.
Evolution-oriented visualizations focused on the training process of the network. 
Re-VACNN~\cite{chung2016re} visualized layer activations in real-time during CNN training. 
CNNComparator~\cite{zeng2017cnncomparator} used matrix visualization to compare neuron features between model snapshots.

\revise{R1-7}{
While the above work significantly addressed the non-transparency issues of deep neural networks, they cannot be directly used for explaining quantum neural network due to the quantum-specific properties.
For example, 
the classical systems cannot reveal how the classical data has been transformed during the quantum encoding procedure because the quantum states are actually complex numbers and cannot be visualized.
}

\subsection{Performance Evaluation of Quantum Data Encoder}

Three prior works investigated the empirical studies for a better characterization of quantum data encoders.
For instance, Rath et al.~\cite{rath2024quantum} conducted a broad empirical analysis of quantum encoding methods across classical machine learning models, highlighting the encoding technologies that improved model accuracy and feature representation efficiency.
\revise{R1-6}{
Monnet et al.~\cite{monnet2024understanding} also used an empirical study to evaluate four differnt encoding strategies, including the angle encoding, in QNN design.
}
Abuga~\cite{abuga2023comparative} focused on different encoding techniques and tested their performance on the MNIST dataset. 
Meanwhile, some works were also conducted to evaluate other aspects of quantum data encoding techniques.
\revise{R1-6}{
Perez et al.~\cite{perez2020data} introduced the concept of data re-uploading, where encoded data is repeatedly fed through quantum layers to address the challenge that quantum states cannot be copied or duplicated (a principle known as the no-cloning theorem).
}
Hubregtsen et al.~\cite{hubregtsen2021evaluation} proposed quantum metrics like expressibility and entanglement capacity to evaluate how different encoders influenced variational circuit outputs.
Lin et al.~\cite{lin2017data} studied encoding strategies tailored for noisy intermediate-scale quantum (NISQ) hardware and balanced circuit complexity with encoding fidelity for practical implementations.
Romero et al.~\cite{romero2017quantum} examined how encoding using Pauli matrices enhanced the separability of quantum states, leading to better feature discrimination in supervised learning tasks.

The prior works mainly focused on leveraging different empirical study approaches (e.g., metric evaluation and comparative study) 
to investigate how to evaluate the quantum data encoders' performance.
\reviseSecond{Despite the popularity, the conventional methods, \textit{e.g.}, the expressibility or end-to-end testing, cannot directly build a mental intuition and further support a better understanding of the encoder by utilizing the human-in-the-loop methods.}
\revise{R1-7}{
Our study focuses on this topic via a design study by collaborating with domain users and experts.
}

\section{Background}

In this section, we introduce some fundamental concepts (i.e., quantum data, QNNs, and quantum data encoders) to help readers better understand the basics of quantum computing and QNNs.

\textbf{Quantum data.}
Quantum data refers to information stored in quantum states, which are the fundamental units of quantum information. Unlike classical data represented by bits that can be either 0 or 1, quantum data is represented by qubits that can exist in a superposition of both 0 and 1 states simultaneously. Mathematically, a quantum state $\ket{\psi}$ of a single qubit can be expressed as a linear combination of the basis states $\ket{0}$ and $\ket{1}$:

\begin{equation}
\label{equation:1}
\ket{\psi} = \alpha\ket{0} + \beta\ket{1},
\end{equation}
where $\alpha$ and $\beta$ are complex numbers such that $|\alpha|^2 + |\beta|^2 = 1$. 
% This property of superposition allows quantum computers to perform computations on multiple states simultaneously, providing a significant advantage over classical computers.
Another fundamental property of quantum data is entanglement, which occurs when the quantum states of two or more qubits become interconnected such that the state of one qubit cannot be described independently of the states of the other qubits. 
Entanglement allows for the creation of correlations between qubits that are not possible in classical systems.

In quantum mechanics, the density matrix is a fundamental mathematical framework used to describe the state of a quantum system.
\revise{R1-5}{
For example, consider a two-qubit system, i.e., a Bell state, which is a well-known pure state in the category of quantum states, to illustrate the concept. The density matrix corresponding to this state is given by:
}

\begin{equation}
\label{equation:4}
\rho = |\Phi^+\rangle \langle \Phi^+| = \frac{1}{2} \begin{pmatrix}
1 & 0 & 0 & 1 \\
0 & 0 & 0 & 0 \\
0 & 0 & 0 & 0 \\
1 & 0 & 0 & 1
\end{pmatrix}.
\end{equation}

This density matrix fully captures the entangled nature of the Bell state, as it encodes the correlations between the two qubits. If the two qubits were in a mixed state, such as a probabilistic combination of \(|00\rangle\) and \(|11\rangle\), the density matrix would instead represent the weighted sum of these states.

\textbf{Quantum neural networks.}
QNNs are neural networks that leverage the principles of quantum computing to perform computations. QNNs consist of three basic components: the encoder, the ansatz, and the measurement. The encoder transforms classical input data into quantum states.
The ansatz, also known as the quantum-version model layer, is a parameterized quantum circuit that applies a series of quantum gates to the encoded quantum states. 
\revise{R1-5}{
The general form of an ansatz can be represented as: $\ket{\psi(\theta)} = U(\theta)\ket{\phi}$, 
where \( U(\theta) \) is a unitary operator parameterized by a set of parameters \( \theta \), and \( \ket{\phi} \) is the encoded quantum state.
} The parameters \( \theta \) are typically trained parameters that are optimized during the training process to minimize a cost function.
Measurement is the process of extracting classical information from quantum states after they have been processed by the ansatz. 
The measurement outcome is a probability distribution over the possible states. Mathematically, the probability of measuring a state \( \ket{x} \) from the quantum state \( \ket{\psi(\theta)} \) is given by:

\begin{equation}
\label{equation:6}
P(x) = |\braket{x|\psi(\theta)}|^2,
\end{equation}
where \( \braket{x|\psi(\theta)} \) is the inner product of the measured state \( \ket{x} \) and the quantum state \( \ket{\psi(\theta)} \).

\textbf{Quantum data encoder.}
The quantum data encoder, also known as quantum state preparation, is a crucial component in QNNs. It transforms classical input data into quantum states that can be processed by the quantum circuit. Essentially, the encoder is formed by a series of quantum gates that map classical data into the quantum state space. Notably, one of the most popular methods today is angle encoding, where the values of classical data determine the rotation angles of quantum gates, such as Pauli rotation gates. This transformation converts a data value into a quantum state representation.

\revise{R1-5}{Intuitively, the role of the quantum data encoder can be compared to the input layer of a classical neural network where input features are scaled and passed into neurons for further computation. Similarly, in a QNN, an encoder maps input features into the quantum state space through quantum gate operations (e.g., rotations), allowing the quantum circuit to process information in superposition, giving the network the ability to evaluate many feature combinations simultaneously.}

Once the data is encoded into a quantum state, it cannot be directly inspected, because any measurement collapses the state and destroys its quantum properties such as superposition and entanglement. The encoded data is then passed to the ansatz for further processing, and only after the optimization process is completed can the final quantum state be measured—ensuring that quantum properties are preserved throughout the computation.
\section{\revise{R1-Minor}{Formative Study}}
\label{sec:formative}

To inform the design of our visualization system \toolName, we worked closely with domain experts and collected their actual needs in routine tasks.
In this section, we first introduce the formative study and report on the design
requirements distilled.

\subsection{\revise{R1-Minor}{Formative Methods}}

Following the core stages of the design study as formulated by  Sedlmair et al.~\cite{sedlmair2012design}, we designed the formative study as follows:

\textbf{Participants.}
The study involved nine domain experts (\textbf{P1}-\textbf{P9}) (3 females $+$ 6 males, $age_{avg} = 33.6$, $age_{sd} = 5.8$) from academic institutions across the United States, Singapore, and Sweden. 
Specifically, \textbf{P1-3} are professors, \textbf{P4-6} Ph.D. students, and \textbf{P7-9} postdoctoral researchers.
All participants have research experience in quantum computing and quantum machine learning, with an average of 6.8 years of experience in the field.
Specifically, \textbf{P1}, \textbf{P3}, \textbf{P6-8} work on Variational Quantum Circuits (VQCs).
\textbf{P4-5} specialize in Quantum Reinforcement Learning,
and \textbf{P2} and \textbf{P9} focuses on Quantum Chemistry and Quantum Federated Learning.
Among them, \textbf{P8} and \textbf{P9} are the collaborators of this research project.

\textbf{Process.}
To develop a visualization framework that effectively addresses the needs of quantum computing domain experts, we adopted a problem-driven methodology inspired by the design study framework. In five months, we engaged in a structured and iterative collaboration with domain experts to identify challenges, gather requirements and refine our solution through continuous feedback and evaluation.

\textit{\reviseSecond{Learning phase:}} 
We conducted one-on-one, semi-structured interviews with all domain experts, aiming to elicit insights about their workflows and challenges regarding the interpretability of quantum encoders. 
During this phase, we focused on understanding the specific problems faced by the experts in analyzing quantum encoders, as well as identifying gaps in existing tools and methods. 
The interviews were complemented by an analysis of the experts’ real tasks and datasets to ground our understanding in their real-world context.

\textit{\reviseSecond{Design phase: }}
We then synthesized the insights gathered from the interviews and task analysis into a set of initial design requirements. Based on these requirements, we developed a low-fidelity prototype aimed at addressing the core interpretability challenges identified during the interviews to provide interpretable insights into quantum encoders.

\textit{\reviseSecond{Evaluation phase:}}
We further engaged the domain experts in iterative expert reviews of the prototype. These reviews were conducted in think-aloud sessions, where experts were encouraged to freely explore the prototype and articulate their thoughts, concerns, and suggestions. 
This feedback was instrumental in identifying usability issues, refining the visualization techniques and ensuring that the tool aligned with the experts’ workflows. Each iteration of the prototype was followed by targeted refinements, addressing specific concerns raised during the reviews while ensuring that the overall design remained consistent with the identified requirements.

\textit{\reviseSecond{Deployment phase: }}
Finally, we delivered a refined version of the visualization framework to the domain experts for extended use in their tasks. This phase allowed us to validate the usability and effectiveness of the framework in addressing the interpretability challenges of quantum circuits. 
Feedback from this phase was incorporated into our discussion, the lessons learned documented to inform future work.

\subsection{Design Requirements}
We collected all feedback from the formative study and compiled five design requirements towards supporting a better understanding of the quantum data encoder.

\begin{itemize}
    
    \item[\textbf{R1}] \textbf{Visualize the encoded feature map.}
    Most experts (\textbf{P2, P4, P7-9}) strongly agreed that it would be helpful to  show the encoded feature map to users, \reviseSecond{which could directly show how the encoder processed the original dataset by the visual comparison.}
    \textbf{P5-6} also suggested the decision boundary between the encoded feature map and the original dataset could highlight how well the encoder distinguishes the different classes.

    \item[\textbf{R2}] \textbf{\revise{R1-Minor}{Support the analysis of the mixture of quantum states.}}
    Five experts (\textbf{P1}, \textbf{P3-5}, \textbf{P7}) expected to observe the encoded data from a ``quantum'' perspective.
    Specifically, they suggested enabling the analysis of how the quantum states with different classes are mixed, which can directly reflect the quality of the encoding process.
    \reviseSecond{The motivation, mentioned by \textbf{P7}, is that the quantum-version mixed states could be more accurate than the measured results from the encoder in some cases (i.e., the encoded feature map mentioned in R1) because the quantum states may vary even though the measured values are the same.}

    \item[\textbf{R3}] \textbf{Unveil the process of how data points are encoded.}
    According to the feedback from four experts (\textbf{P1-2}, \textbf{P5-6}), it will be helpful to reflect the process \revise{R2-4}{of} data encoding, which can inform the users of the functionality of each quantum gate in a data encoder.
    Moreover, \textbf{P2} commented that \textit{``The explanation of the intermediate state in each step means everything for the developers.''}

    \item[\textbf{R4}] \textbf{Enable flexible trials with different encoders and datasets.}
    \reviseSecond{To better support the understanding process,} five experts (\textbf{P2}, \textbf{P4-6}, \textbf{P9}) suggested that the visualization system should not be limited to a fixed example and instead enable flexible experiments with different data encoders along with various datasets.
    \textbf{P5} emphasized that QNN users can only gain the key knowledge of how the encoders capture the features by back-and-forth comparisons.

    \item[\textbf{R5}] \textbf{\revise{R1-Minor}{Bridge the gap with the classical neural network.} }
    Five experts (\textbf{P2-3}, \textbf{P5-7}) appreciated systems that can integrate the conventional visualizations for model performance tracing, e.g., the line charts for visualizing the accuracy changes across epochs.
    Meanwhile, they all pointed out the importance of supporting the fine-tuning of the hyperparameters (e.g., learning rates), \reviseSecond{which can connect the dots between the training of classical neural networks.}
     \textbf{P2} also encouraged us to enable the observation of the trained feature map after each epoch of the training process has been completed, \reviseSecond{which could support  a direct analysis of the training for each data point. }

\end{itemize}

\section{\revise{R1-8}{Datasets}}

\revise{R3-3}{
In this paper, we use 2-dimensional data as examples to illustrate the effectiveness of the proposed visualization tool. 
}
As depicted in Figure~\ref{fig:teaser}\component{C}, each cell in the grid represents an individual data point. 
The x and y axes correspond to the two features of the dataset, 
while the color of each cell denotes the label of the respective data point. 
To process this dataset within a quantum neural network framework, we utilized the 2-qubit QNNs, 
which \revise{R2-4}{consist} of the following three components:
1) \textit{Encoder}: transforming the 2-dimensional classical data points into quantum states, which can be represented by a $4 \times 4$ matrix;
2) \textit{Ansatz}: A parameterized quantum circuit that applies single-qubit Pauli rotation gates to the above encoded states. 
For all examples integrated in our visualization system, we fixed the ansatz to support the analysis of the encoder's pattern by implementing different encoders.

\section{\revise{R1-Minor}{\toolName}}

We proposed \toolName, a visualization tool to facilitate the understanding of how the quantum encoder affects the training of QNNs.
In this section, we first introduce the \textit{Encoder Expectation Measurement} to simulate the encoded states before they are fed into the training model.
Then, we illustrate the developed visualization system \toolName{} together with the justification of the system design.
\revise{R1-10}{Note that \textit{Encoder Expectation Measurement} and the visualization system are based on the simulator provided by the \textit{Pennylane}~\cite{pennylane} framework.}

\subsection{Encoder Expectation Measurement}

In QNNs, directly measuring the quantum states after the encoder layer would immediately terminate the training process, because any measurements will collapse the quantum states, leaving no valid information for subsequent layers (i.e., the ansatz) in the next epoch. 
To address this challenge and obtain the encoded quantum states between the encoder and ansatz layers, we propose a novel approach \textit{Encoder Expectation Measurement} to calculate the encoder expectation values.

\begin{figure}[t]% specify a combination of t, b, p, or h for top, bottom, on its own page, or here
  \centering % avoid the use of \begin{center}...\end{center} and use \centering instead (more compact)
  \includegraphics[width=\columnwidth
  ]{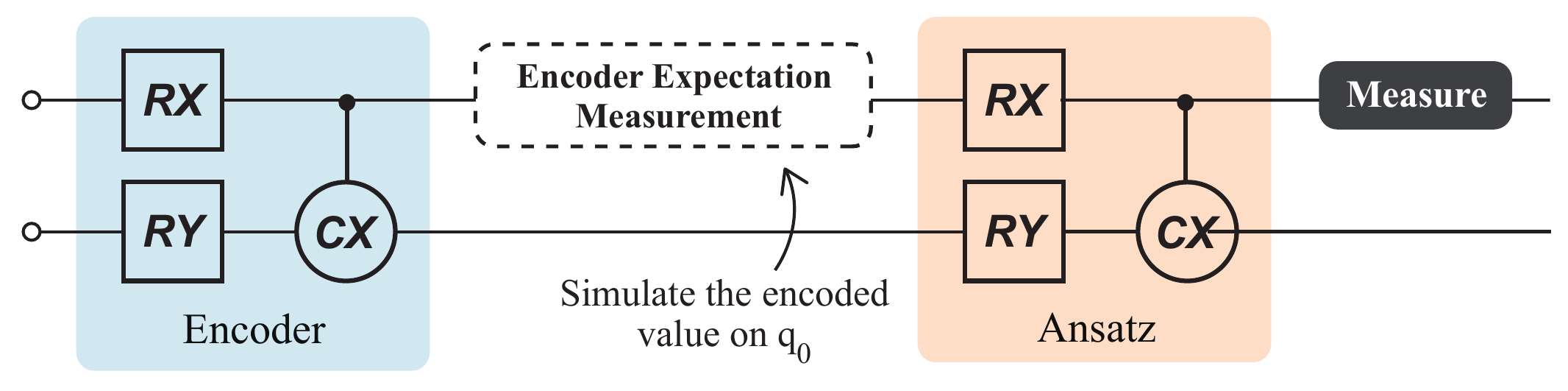}
  \caption{
The illustration of the Encoder Expectation Measurement simulation without directly implementing the measurement gate.
We simulate the encoded states on the first qubit, aligning with the configuration of the actual measurement of $q_0$.
  }
  \label{fig:2}
\end{figure}

\revise{R1-10}{
To achieve this, we first extract the quantum system states by setting the flag after the encoder layer ends,
which provides the density matrix of each quantum state, as shown in Equation \ref{equation:4}.
}
This step allows us to capture the encoded quantum state without directly measuring it. 
As shown in Figure~\ref{fig:2}, since the first qubit (\(q_0\)) is measured as the output of the quantum circuit in every training epoch, the probabilities of \(q_0\) being in the states \(|0\rangle\) and \(|1\rangle\) are calculated, denoted as \(\text{Prob}(q_0 = 0)\) and \(\text{Prob}(q_0 = 1)\), respectively. 
So, the expectation value could be $\text{Prob}(q_0 = 0) - \text{Prob}(q_0 = 1)$.
Given the method of single-qubit probability calculation, the expectation value of the encoded quantum state at the encoder output can be computed as:

\begin{figure}[t]% specify a combination of t, b, p, or h for top, bottom, on its own page, or here
  \centering % avoid the use of \begin{center}...\end{center} and use \centering instead (more compact)
  \includegraphics[width=\columnwidth
  ]{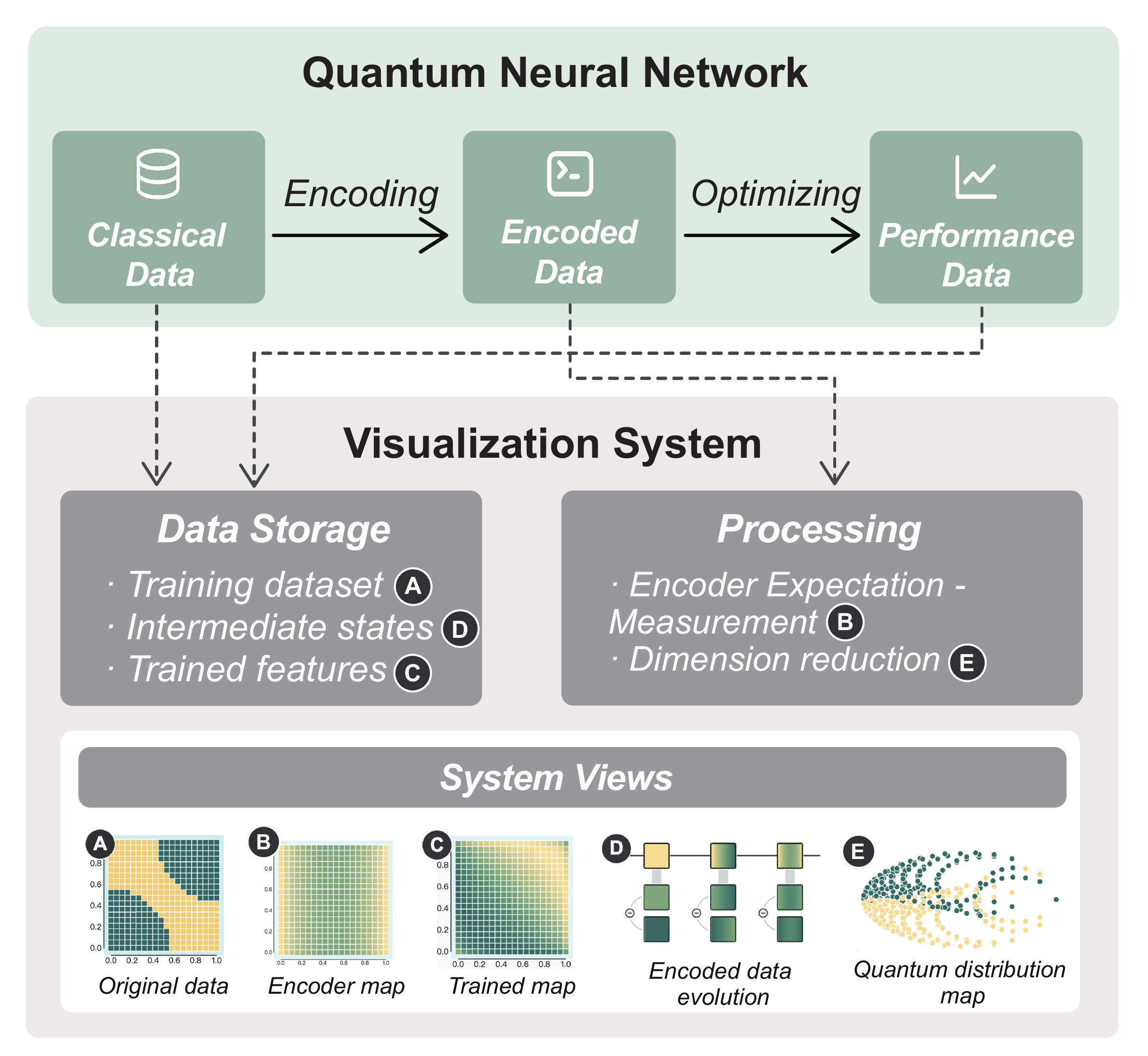}
  \caption{
The system of \toolName{} contains three modules: a Data Storage module, a Processing module, and a System Views module.
The Encoded Data will be extracted from the QNN training process and then delivered to the Processing module for the generation of the two views, i.e., the Encoder map and the \revise{R1-Minor}{State Comparison Map}.
  }
  \label{fig:4}
\end{figure}

\begin{equation}
\label{equation:8}
E_{\text{encoder}} = \sum_{i \in \{0,1\}} \mathrm{Pr}(\ket{0i}) - \sum_{j \in \{0,1\}} \mathrm{Pr}(\ket{1j}),
\end{equation}
where \(E_{\text{encoder}}\) represents the encoded value of the corresponding classical data point. This expectation value is then used as the encoded representation of the classical data point, ensuring that the encoded quantum state is preserved and passed to the ansatz layer without collapsing the quantum circuit.

\begin{figure*}[t]% specify a combination of t, b, p, or h for top, bottom, on its own page, or here
  \centering % avoid the use of \begin{center}...\end{center} and use \centering instead (more compact)
  \includegraphics[width=\linewidth
  ]{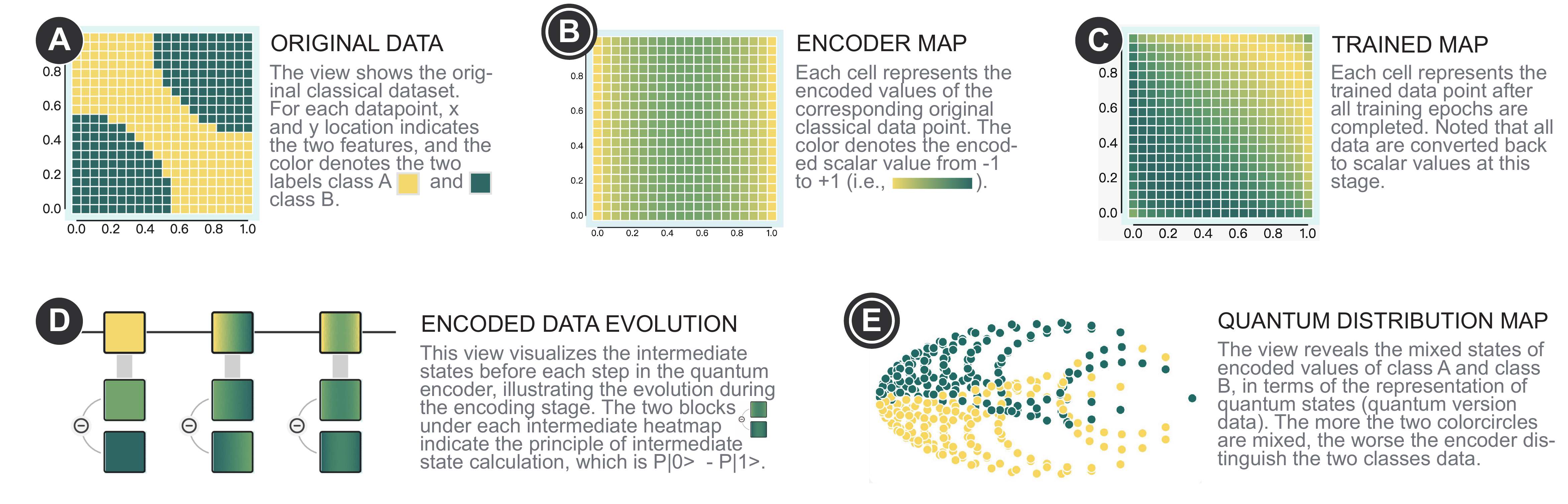}
  \caption{
All views implemented in \toolName{}.
(A) The visualization of the original 2-dimensional training dataset.
(B) The Encoded Map view supports the direct inspection of the encoded classical data, where each cell represents the expectation value of the quantum states.
(C) The Trained Map view shows the features of the trained dataset after the training process is finished.
(D) The Encoded Data Evolution view shows the evolution before the all data points are completely encoded, where each group of heatmap indicates the states before each step in the encoding process.
(E) The \revise{R1-Minor}{State Comparison Map} view denotes the mixed states of the two-class data points, assessing the encoder's performance to distinguish between the features of two-class data points.
Note that the work shown in sub-figures (B) and (E) are the key contributions of our paper.
  }
  \label{fig:3}
\end{figure*}

\subsection{\revise{R1-Minor}{Visual Interface}}

In this subsection, we illustrate the visualization system \toolName{} in terms of the implemented views, followed by the user interactions and design justifications.
Figure~\ref{fig:4} shows the architecture of \toolName{} and how it integrates with the external QNNs.

\subsubsection{Classical Data Representation}

To provide users with an intuitive understanding of the original data and facilitate comparisons with the encoded quantum states, 
we developed the ``Original Dataset View''.
Shown in Figure~\ref{fig:3}\component{A}, this view visualizes the original two-dimensional input before it is encoded into quantum states. 
Specifically, each data point is represented as a cell, where the horizontal and vertical positions correspond to the two dimensions of the dataset. The data points are color-coded to indicate their respective classes.
To standardize the encoding process, we assign numerical values of \(-1\) and \(+1\) to the two classes, respectively. 
Note that for all datasets implemented in the system, the horizontal and vertical features are evenly distributed within the range of 0 to 1.

\subsubsection{Encoded Data Representation}

We present the Encoder Map View to provide users with a direct visualization of the quantum states after the encoding process, enabling them to observe and analyze the patterns generated by the encoder. 
As described in Section~\ref{fig:2}, we leverage the Encoder Expectation Measurement to compute the encoded quantum states by mapping it back to a scalar value. 
This scalar value serves as a bridge between the complex quantum states and an interpretable visual representation, allowing users to intuitively understand the encoding process.
Then, based on the scalar values, the Encoded Map View visualizes all encoded data points in a heatmap-like view, as shown in Figure~\ref{fig:3}\component{B}. 
The principles of the visual design of the above Encoded Map are consistent with the Original Dataset View.
The key difference lies in the color encoding, where the color of each cell in the Encoded Map represents the expectation value of the encoded quantum states for the corresponding data point, ranging from \(-1\) to \(1\).

To further support the rationale behind the Encoded Map, we developed a visualization called the Encoded Data Evolution View, as shown in Figure~\ref{fig:3}\component{D}. 
It illustrates how the encoded map is generated during the encoding process before the encoder is finalized. 
The first row of heatmaps in this view depicts the intermediate states of the encoder at each step of the encoding process, providing users with insights into how the encoding evolves over time.
Moreover, to explain the intermediate quantum states in greater detail, the second and third rows of the Encoded Data Evolution View visualize the probabilities of qubit \(q_0\) being in states \(|0\rangle\) and \(|1\rangle\), respectively, aligning with Equation~\ref{equation:8}.

\subsubsection{\revise{R1-Minor}{State Comparison Map}}

To better portray the encoded data, we presented another visualization called ``\revise{R1-Minor}{State Comparison Map}'' (Figure~\ref{fig:3}\component{E}), a two-dimensional scatter plot designed to illustrate how well the different-label data points are encoded, 
which can intuitively reflect encoder's ability to distinguish between different classes of data.

The system captures intermediate quantum states at designated points, which are placed at key stages.
% , such as after the encoder steps.
For each data point, the extracted state provides a complex state vector, which is subsequently converted into a density matrix representation. 
Given that even a small number of qubits can result in high-dimensional density matrices, i.e., the density matrix with the size of $2^N \times 2^N$ for $N$-qubit case, as illustrated in Equation~\ref{equation:4}.
\revise{R1-11}{
We then employ dimensionality reduction techniques to create a more interpretable representation. The density matrix is first flattened into a single numeric array by Principal Component Analysis (PCA) to reduce these high-dimensional arrays into 2D coordinates. 
}
\revise{R2-2}{
Unlike nonlinear methods such as t-SNE, we chose PCA because it provides a deterministic projection that preserves the global variance structure of the encoded quantum states and also allows a direct linear mapping from the high-dimensional density matrix to two principal axes, aligning with the system’s goal of enabling transparent and reproducible visual reasoning. 
}
PCA identifies the directions (i.e., principal components) along which the data exhibits the greatest variance and projects each flattened density matrix onto these components. 
This processing stage can \revise{R3-5}{capture} the most significant features of the quantum states while ensuring the data remains visually interpretable. 
Each projected sample is paired with its original class label (e.g., \(+1\) or \(-1\)).
% , enabling graphical distinctions such as color-coding in the subsequent visualization step.
Once the PCA-generated coordinates are paired with their class labels, the system visualizes the data as a two-dimensional scatter plot.
As shown in Figure~\ref{fig:3}\component{E}, each data point is positioned according to its corresponding PCA coordinates, with its class label determining its color, allowing users to visually assess the separability of different classes that varies significantly depending on the encoder design.

\begin{figure}[!h]% specify a combination of t, b, p, or h for top, bottom, on its own page, or here
  \centering % avoid the use of \begin{center}...\end{center} and use \centering instead (more compact)
  \includegraphics[width=\columnwidth
  ]{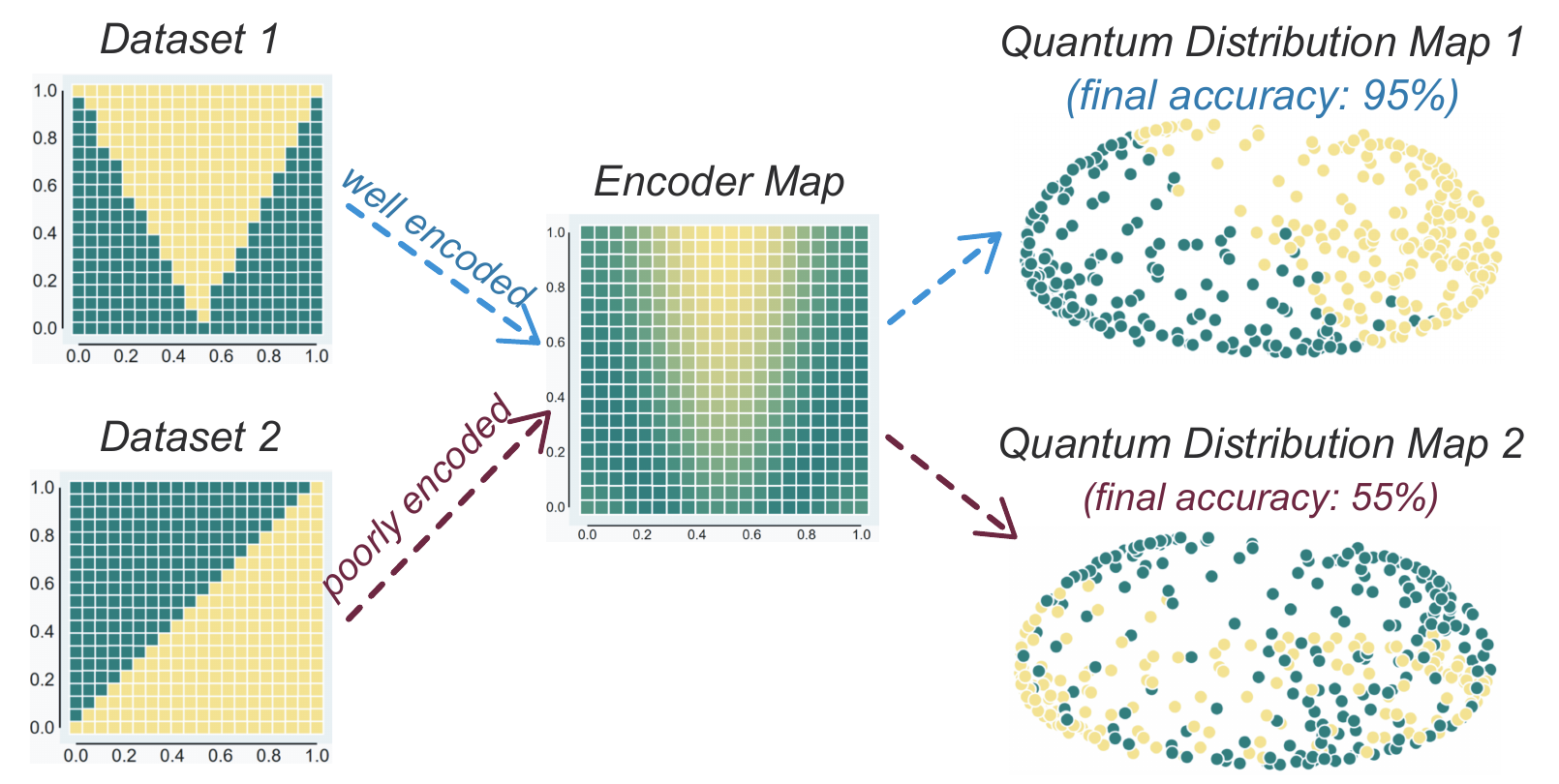}
  \caption{
The \revise{R1-Minor}{State Comparison Map} compares two scenarios. 
Top: a well-encoded scenario with an accuracy of 95\%, where distinct clusters corresponding to different classes are clearly visible;
Bottom: a poorly-encoded scenario with an accuracy of 55\%, where overlapping clusters indicate insufficient class differentiation. 
  }
  \label{fig:5}
\end{figure}

For example, as indicated by Dataset 1 and \revise{R1-Minor}{State Comparison Map} 1 in Figure~\ref{fig:5}, when the encoder effectively separates the classes, the scatter plot displays distinct clusters, with each cluster corresponding to a specific class. 
Such clear separation indicates that the encoding process preserves meaningful feature distinctions in the quantum state space, leading to a decent accuracy of $95\%$.
Conversely, for the poorly-encoded data like Dataset 2 in Figure~\ref{fig:5}, if the scatter plot shows extensive mixed patterns between multiple classes, the encoder fails to promote sufficient differentiation.
In such cases, the prediction accuracy is usually very low ($55\%$ in this case) due to the mismatch of the original dataset and corresponding Encoder Map.

\subsubsection{Model Performance Analysis}

To bridge the gap between quantum and classical neural networks training, we implemented a widely used approach to visualize the changes in performance throughout the training process. 
As shown in Figure~\ref{fig:teaser}\component{H}, we utilized line charts to depict the training loss and accuracy across each epoch of the QNN. 
Specifically, the blue line represents the training loss, while the orange line illustrates the accuracy. 
Furthermore, to better explain the accuracy, we developed another view to directly visualize the trained dataset and the resulting classification. 
As shown in Figure~\ref{fig:3}\component{C}, this view presents the learned patterns after the training process is finalized. Each cell in this visualization corresponds to a data point, with its color representing the predicted value for that point. This design allows users to compare the original dataset with the trained dataset, offering a direct understanding of how well the model has captured the underlying patterns.

\subsubsection{User Interactions}

We also developed various user interactions to support the flexible exploration of \toolName{} in terms of the QNN model training and the pattern comparison of original data and encoded data.

\textbf{Encoder selection. }
To enable users to explore a wide range of encoder options, we implemented an encoder selection feature. 
As shown in Figure~\ref{fig:teaser}\component{A}, users can click the arrows in the component named ``Specify Encoder'' to open a side-selection panel, allowing users to freely select from a set of pre-configured encoder circuits.
Specifically, we prepared a total of 10 2-qubit encoder circuits, incorporating various combinations of Pauli Rotation Gates and Control Gates.

\textbf{Model configuration.}  
Users can also control the training process through a control panel. 
For example, the panel allows users to configure key hyperparameters such as the number of training iterations and the learning rate. 
These parameters can be adjusted either by directly inputting the desired values or by dragging the slider handles to fine-tune the settings. 
Additionally, the control panel includes buttons that enable users to pause or resume the training process at any time, offering greater flexibility in managing the execution of the QNN model. 
This interactive control mechanism ensures that users can experiment with different training setups and transparentize their effects on the QNN model's behavior.

\subsubsection{Design Justification}

\revise{R3-4}{
Based on the interactive test of the deployment with the participants during the formative study,
we carefully refined \toolName{} by receiving valuable feedback.
Below, we outline the key design decisions and their justifications:
}

\textbf{Cell shape. }  
Initially, we used circular shapes to represent each data point in the heatmap. 
However, the gaps between the circles made it difficult to inspect the overall patterns, as the empty spaces diluted the visual continuity of the heatmap. 
To mitigate this issue, we switched to square cells, which can seamlessly tile without gaps, \revise{R3-5}{allowing a clearer comparison of the two colors representing different classes.}

\textbf{Color distribution. }  
\revise{R3-4}{
We initially employed a color range with three key points (e.g., \(-1\) \iconRange \(+1\)), where white was used as the middle point,
but this approach caused the colors near the white midpoint to appear too faint.
In response, we changed the color distribution to use a two-keypoint range by eliminating the white midpoint. 
}

\textbf{Circuit display. }  
\revise{R3-4}{
Feedback from the experts highlighted the importance of aligning the visualization of encoded data with the actual encoder circuit used, 
which provides a direct reference to the encoder's structure, so we implemented the encoder circuit diagram in \toolName.
}

\textbf{Color palette selection.}  
Since the encoded data generated by the Encoder Expectation Measurement often consists of decimal values, the palette needed to clearly differentiate between the two classes while preserving subtle variations in the data. 
We experimented with a number of palettes (candidates: \iconOptionA, \iconOptionB, \iconOptionC, \iconOptionD, \iconOptionE) 
and ultimately selected the current one, which strikes a balance between highlighting the encoded patterns and ensuring that decimal values are easily distinguishable as belonging to Class A or B.

\revise{R1-12}{
\subsubsection{Implementation}
\toolName\ integrates a \textit{React}-based frontend with a \textit{Flask} backend to enable interactive visualization of quantum encoder performance. 
The backend provides both synchronous circuit evaluation and asynchronous streaming training via a session-based API, supporting real-time pause, resume, and stop controls. 
Server-sent events (SSE) are employed to deliver incremental updates of model performance and quantum state distributions to the frontend, facilitating progressive visualization of training dynamics. 
}
\section{Evaluation}

\revise{R1-13}{
We conducted a two-step process to evaluate the proposed visualization system, which consists of two case studies to discover the findings and a post-study interview for qualitative feedback.
We conducted the case studies in a think-aloud manner, where participants were asked to freely explore the system while verbalizing their reasoning and observations. 
During the exploration, we carefully observed their interactions with the interface and recorded screen activities and spoken reflections. 
Following the system use, we carried out semi-structured interviews to gather qualitative feedback. The interview protocol focused on four aspects—effectiveness, usability, workflow, and visual design—using a set of guiding questions (e.g., “How effective do you find the system in helping you assess encoder quality?”, “Does the workflow align with your normal analysis process?”, or “How do the visualizations support or hinder your reasoning?”). 
The overall research goal was to examine how well the system supported expert reasoning about encoder quality and to identify design opportunities for improving explainability of encoders.
}
To control variables and isolate the effect of the encoder, we implemented the same ansatz across all examples.
All the invited experts in this stage are different from the participants involved in the stage of Design Formation illustrated in Section~\ref{sec:formative}.

\subsection{Case Study 1: Reasoning Model Failures Through the Encoder Capability}

In this case study, we collaborated with \textbf{P1}, a post-doctoral researcher from Sweden who has 8 years of expertise in QNNs and quantum federated learning. 
We selected this expert because of his deep knowledge of the domain. 
The goal of the study was to have the expert explore the system, identify an optimal encoder for a given dataset, and formulate the rationale behind his findings.

To begin, \textbf{P1} examined the six open dataset options available in the Dataset Selector View (Figure~\ref{fig:teaser}\component{B}). 
After a brief glance, he selected the third dataset, which features a circular pattern in the top-right corner.  
The Original Dataset View (Figure~\ref{fig:teaser}\component{C}) is updated accordingly, displaying the selected dataset.  
\textbf{P1} remarked, \textit{``This dataset has a clear pattern with the completed circle boundary, so I expect the encoded data to preserve the same clear patterns, with the class-B data clustering tightly together.''}
Next, he used the Circuit Selection Component (Figure~\ref{fig:teaser}\component{A}) to select the encoder configuration \texttt{RX-RY-RY-CNOT}. 
He explained, \textit{``This encoder seems to have stronger expressibility and entanglement capability. I think this encoder should be sufficient to encode this 2D dataset well.''}

To evaluate the encoder’s performance, \textbf{P1} then launched the training process by clicking the ``Play'' button in the center of the control panel. 
The training process began, with the Trained Map and Performance View dynamically updating as the epochs progressed.  
Upon the completion of training, \textbf{P1} turned his attention to the Encoder Map View, which displayed the encoded values using the Encoder Expectation Measurement method.  
Then, the Encoder Map View is updated to show the encoded data distribution, as shown in Figure~\ref{fig:teaser}\component{F}.  
He commented, \textit{``This is surprising because I’ve never seen what the encoded quantum data looks like before.''}
\textbf{P1} then carefully examined the Encoder Map View to analyze the patterns in the encoded data.  
The view revealed unexpected artifacts, including two stripe-like patterns, which were not present in the original dataset.  
\textbf{P1} remarked that from this view he does not think this encoder fits the dataset, since the patterns in the original dataset are completely lost, and he had no idea why there are two stripe-like patterns in the Encoder Map.

\textbf{P1} then used the Performance View (Figure~\ref{fig:teaser}\component{H}) and the Trained Map (Figure~\ref{fig:teaser}\component{G}) to evaluate the model’s overall performance.  
The Performance View showed the model accuracy performance of 55\%, while the Trained Map revealed that the model had learned an incorrect pattern, with class-A patterns dominating the four corners of the map.  
He noted, \textit{``The accuracy is pretty bad, especially since the ansatz is good enough for this simple 2D dataset. The heatmap explains why—somehow, the model learned a pattern where the four corners are class A, which I guess is caused by the bad encoding.''}

To gain further insights into the encoding process, \textbf{P1} further inspected the Quantum Distribution Map View as hinted by us.  
As shown in Figure~\ref{fig:teaser}\component{J}, the view revealed that data points from different classes were mixed together, with no clear separation of the circular pattern.  
He commented, \textit{``This distribution  exactly explains why the selected encoder has unsatisfactory accuracy performance. Since the encoder completely mixes the quantum states from different classes, the circular pattern of class-B data cannot be classified at all in this view.''}
Finally, the expert analyzed the Encoded Data Evolution View (Figure~\ref{fig:teaser}\component{E}) to understand how the encoder evolved during the encoding process.  
The view showed that, although the gates were expressive enough, the linear angle preparation was insufficient to capture the circular pattern in the dataset.  
\textbf{P1} concluded, \textit{``Now it is clear that the gates are strong enough, but the linear angle preparation cannot fit the circular pattern for this dataset.''}

After a follow-up open discussion and reflecting on the entire analysis, \textbf{P1} summarized his findings.  
He said, \textit{``I have a hypothesis in mind. The Encoded Map View shows how closely the encoded pattern matches the original dataset, which can be inferred from the convergence speed of model performance. Meanwhile, the Quantum Distribution View can reflect the encoder’s ability to recognize patterns, which can be seen from the fluctuations during the convergence process.''}

\begin{figure*}[t]% specify a combination of t, b, p, or h for top, bottom, on its own page, or here
  \centering % avoid the use of \begin{center}...\end{center} and use \centering instead (more compact)
  \includegraphics[width=\linewidth
  ]{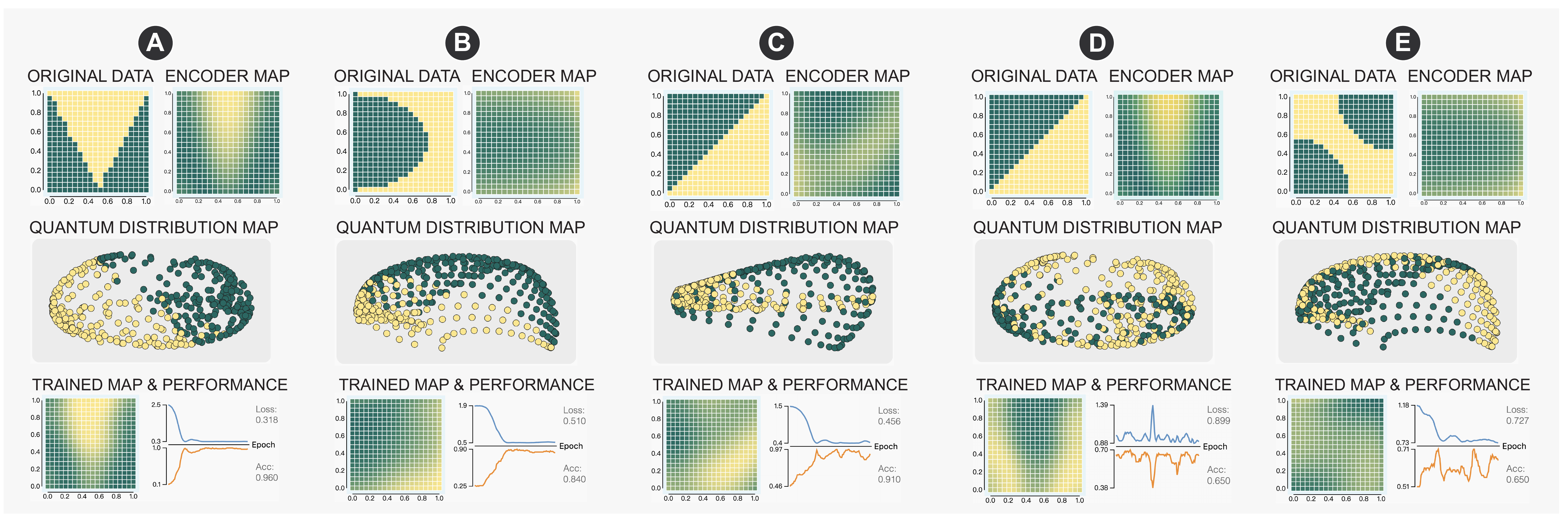}
  \caption{
  The results in different usage scenarios, with the encoding performance varying from well-coded to poorly-encoded levels.
  (A)(B) The Encoder Map and Quantum Distribution Map show the encoder encodes the original dataset well, resulting in good accuracy.
  (C) The Encoder Map shows the encoder mapped the features close to the original datasets, but the Quantum Distribution Map shows the encoder still cannot separate different classes apart.
  (D)(E) The Encoder Map and Quantum Distribution Map perform badly in encoding the original datasets, both leading to bad accuracy.
  }
  \label{fig:6}
\end{figure*}

\subsection{Case Study 2: Unveiling Encoder Limitations Through Pattern Preservation and Feature Mapping}

We worked with a professor \textbf{P2} and a post-doctoral researcher \textbf{P3} from the United States to operate our visualization system.
Both experts specialize in quantum systems and applications based on QNNs. 
This case study was conducted online via Zoom.

Unlike the expert \textbf{P1} in Case Study 1, these two participants formed the same hypothesis as \textbf{P1} after a brief introduction to the system, without interacting with it initially.
Therefore, they expected to validate the shared hypothesis through free exploration of the system.

Bearing this idea in mind, the experts first selected the encoder shown in Figure~\ref{fig:6}\component{A}. 
They noticed that the encoded states \revise{R1-Minor}{look} very similar to the sixth dataset. 
So, they chose this combination to validate their hypothesis.  
After launching the training and inspecting the Quantum Distribution Map, \textbf{P2} reported: 
\textit{``The Encoded Map view shows the encoder is good enough for this dataset, which reduces the training burden for the ansatz because this encoder has successfully simplified the classification task. 
The Quantum Distribution Map confirms this, as the two data points are well-separated, showing that the encoding process effectively captured the features of the two classes. 
As a result, the model accuracy shows excellent performance achieving 96\%, with a fast convergence curve and minimal fluctuation.''}

The participants continued their exploration by selecting another dataset. 
As shown in Figure~\ref{fig:6}\component{B}, they found that the Encoder Map closely resembled the dataset.
However, the Quantum Distribution Map was not as good as in the previous example, with some overlapping and disorganized data points at the boundaries. After starting training, the results showed that, although the accuracy was still decent, the convergence process was longer compared to Scenario~\component{A}.  
\textbf{P2} commented: \textit{``This roughly confirms our hypothesis, but I would like to observe the behavior of some other bad scenarios.''}

Next, they switched to the dataset shown in Figure~\ref{fig:6}\component{C}. This scenario was similar to Scenario \component{B}, 
but the Quantum Distribution Map was more messy, with many data points clustered around the middle horizontal region.  
\textbf{P3} reported: \textit{``That's why the line is more fluctuated, that's very interesting.''}

Curious about the performance when the dataset and encoder map were completely different,  as shown in Figure~\ref{fig:6}\component{D}, they kept the dataset from Scenario~\component{C} but switched to the encoder used in Scenario~\component{A}.  
After the comparison, they observed that under this scenario, with a poor Encoder Map and a messy Quantum Distribution Map, the model struggled to converge. 
The accuracy curve showed significant fluctuations, and even after 100 epochs, the accuracy performance only reached approximately 65\%.  
A similar pattern was observed in Figure~\ref{fig:6}\component{E}, where a poor encoder and distribution map resulted in a prolonged convergence process and a highly fluctuating curve.

\textbf{Conclusion.} 
The experts were excited about the findings in the exploration.
After a short discussion, they concluded that, while the visualization system was highly effective, it also validated two key findings in their research. They summarized the practice as follows:  
If the objective is to optimize QNN model performance, two strategies can be considered regarding the quantum encoder. The first strategy involves designing the encoder to make feature representations closely aligned with the characteristics of the training dataset, thereby reducing the complexity of the subsequent optimization task—effectively shortening the optimization trajectory. The second strategy, as shown in the Quantum Distribution Map, is to encode data points such that features from different classes become clearly distinguishable, thereby resulting in smoother convergence during model training. However, the ultimate optimal model performance still relies on the combined effect of the quantum encoder and the ansatz, but the ansatz design would be a different story, which is beyond the scope of this work.

\subsection{Expert Interview}

We conducted post-study interviews with the domain experts to gather qualitative feedback on the effectiveness, usability, workflow and visual design of \toolName{}.

\textbf{Effectiveness.}  
\revise{R1-14}{The experts were asked about how effective \toolName{} is on understanding the correlation between encoders and model performance, aligning with the design requirements \textbf{R1-4}.}
The experts provide positive feedback regarding the system’s ability to uncover hidden insights about the quality of encoders, which was barely studied in the previous quantum machine learning area.  
\textbf{P2} highlighted, \textit{``From the perspective of our findings, I believe \toolName{} does an excellent job of revealing hidden insights about the encoder's quality,''}
because previously, developers of VQCs were essentially performing blind executions, but this system truly ``visualizes'' the diagnostic capabilities of encoders.
\textbf{P1} emphasized the innovation of the Encoder Expectation Measurement within the quantum computing field, stating, \textit{``From a quantum computing perspective, Encoder Expectation Measurement itself is a significant novel approach in the area of explainable quantum neural networks. 
Since explainability in QNNs is still an under-explored research direction, this contribution is impactful.''}  
\textbf{P3} also praised the Quantum Distribution Map, commenting, \textit{``I’m still surprised by the intuitive visualization of the Quantum Distribution Map. This design makes the encoder’s performance immediately clear, showing how far the encoded states are from the maximum mixed states based on the distribution of data points. This is absolutely critical.''}

\textbf{Usability.}  
The system’s usability was highly regarded, especially for domain experts without a visualization background.  
\textbf{P3} remarked, \textit{``In general, this visualization system is very user-friendly for domain experts without a visualization background. We often use heatmaps to analyze performance, so the heatmaps in this system are simple and practical for us.
Compared to other tailored visualization systems we have used before, they often confused us, while
\toolName{} feels like a very practical dashboard. I could figure out how to use it even without a system introduction.''}
\textbf{P2} appreciated the system’s support for free exploration, noting that the system allows users to freely explore and compare various encoder templates and datasets, enabling a comprehensive understanding in encoders.
Additionally, \textbf{P2} praised the clear tutorial at the bottom of the interface, stating that the tutorial provides a concise introduction to the interface and includes helpful explanations of QNNs and encoders, making it especially friendly for novice students.

\textbf{Workflow.}  
The system’s workflow was described as intuitive and logical, guiding users seamlessly through the analysis process.  
\textbf{P1} commented that the system’s layout and transitions effectively guide user interactions, allowing users to understand how to interact with the system without needing a tutorial. 
He also noted that the current workflow—starting with training and then analyzing the encoder’s quality and performance through the Encoder Map and Quantum Distribution Map—is both reasonable and effective. Furthermore, \textbf{P1} suggested an enhancement that the Quantum Distribution Map could be used to pre-test encoders that perform well for specific types of datasets. Such encoders could then serve as templates for similar datasets in the future.

\textbf{Visual Design and Interactions.}  
The visual design and interactions of \toolName{} received positive feedback from the experts.  
\textbf{P2} commended the intuitive representation of the Quantum Distribution Map, saying that the Quantum Distribution Map clearly visualizes the degree of confusion in the encoded states. 
Additionally, the design and color choices for all the heatmaps are excellent, making it easy to analyze encoded and learned patterns.
\textbf{P3} also highlighted the value of the Encoded Data Evolution view and reported that it provides an in-depth explanation of how different gates impact the dataset from the perspective of gate implementation. 
It is incredibly helpful for understanding the role of different rotation gates and offers valuable knowledge for designing encoders in the future.
Finally, \textbf{P3} praised the smooth interactions within the system, particularly during the training process, \textit{``The interactions are very smooth, especially the ability to pause training at any epoch to inspect the intermediate states. This functionality is extremely practical.''}

\textbf{Suggestions.}
Besides the positive feedback, several
participants also offered constructive suggestions.
First, experts emphasized the need to involve a larger community of researchers to contribute to the system. This expansion could facilitate the identification and design of new typical encoders tailored for datasets with specific features. 
By integrating more diverse encoders, the system could broaden the community’s knowledge base, leveraging intuitive visual representations to drive further advancements.
Second, to support deeper analysis, the system could allow users to trace specific data points of interest. For example, incorporating brushing interactions in the Trained Map View would enable backward-analysis for poorly-encoded or mixed data points.

\section{Discussion and Future Work}

We further discuss four aspects in our research.

\textit{\reviseSecond{Explainable Quantum Artificial Intelligence.}}
Explainable artificial intelligence (XAI) has emerged as an active area of research in classical machine learning, aiming to make AI systems more transparent, interpretable, and trustworthy. 
In the context of quantum artificial intelligence, the need for explainability is even more pressing due to the inherent complexity and non-intuitive nature of quantum mechanics. 
According to our expert interview, they all agreed that this paper represents a foundational step toward the emerging field of XQAI by introducing a method to calculate and visualize the encoder expectation value.
% This method provides an interpretable representation of quantum states.
The methodology proposed in this paper aligns with the goals of XQAI by addressing the ``black-box'' nature of quantum encoders. 
In the future, as the field of quantum computing continues to advance, XQAI will play an important role in ensuring that quantum systems are not only powerful but also interpretable.

\textit{\reviseSecond{Potential Impacts.}}
The proposed approach for calculating and visualizing the encoder expectation value demonstrates significant potential for advancing the understanding and usability of QNNs across multiple domains. 
First, the approach outlined in this paper has the potential in educational scenarios, particularly for lecturers and instructors aiming to teach quantum computing and quantum neural networks. 
Second, for novice developers working on quantum machine learning models, the proposed approach offers an interpretable method for understanding the encoders. 
% This understanding enables them to make informed decisions when designing or optimizing QNNs for specific tasks. 
Furthermore, the ability to visualize the encoded states helps identify potential issues or inefficiencies in the encoding process.
Furthermore, the proposed method also has implications for quantum circuit designers. 
By providing an intuitive understanding of the encoder's performance, this approach facilitates the identification of optimal encoding strategies for specific applications. 
Additionally, the visualization techniques introduced in this paper provide a novel way to analyze and refine quantum circuits, helping designers intuitively explore the relationship between the encoder and model performance.

\textit{\reviseSecond{Scalability and Generalizability.}} 
In this paper, we demonstrated the proposed method using a two-dimensional dataset example, which allowed us to visualize the quantum states in a two-dimensional heatmap-like representation.
\revise{R1-4}{
Our paper focuses on using the 2-dimensional dataset to build a cognitive model for QNN developers, because 2D datasets are straightforward and users can understand without much time and efforts, which is suitable for enhancing the explainability of quantum encoders.
\reviseSecond{However, for more complicated real-world datasets, our system, especially our major contributions ``Encoded Data Representation'' and ``State Comparison Map'', can be generalized by incorporating dimensionality reduction techniques, which can reduce the high-dimensional quantum states into a lower-dimensional representation, which can suit \toolName{} while preserving the essential features of the data. }
}
Furthermore,  
while the primary focus of this work is on analyzing and interpreting quantum encoders, the proposed methodology has the potential to be generalized to the optimization process of QNNs, e.g., ansatz layers. 
For example, the presence of barren plateaus and vanishing gradients in quantum optimization could be better solved by the proposed method.

\textit{\reviseSecond{Future Work.}}
The current system operates on static datasets and pre-defined quantum circuits.
Future work could focus on integrating adaptive learning algorithms into the system, allowing it to dynamically adjust its parameters in response to changes in the input data or circuit configurations.
% Additionally, adaptive learning could improve the system’s performance by optimizing the encoding and circuit parameters iteratively, ensuring that it remains effective across a wide range of tasks and scenarios.
Moreover, the system is tailored for specific tasks at the moment, such as analyzing quantum encoders. Future work could extend its functionality by enabling it to handle diverse quantum machine learning tasks, such as classification, regression, and clustering, within a single framework. Additionally, future work can be further applied to quantum federated learning \cite{ren2023towards
} within a distributed framework.

\section{Conclusion}

In this work, we identified key challenges for improving the explainability of quantum encoders.
Following the distilled design requirements collected from the formative study, we developed \toolName, a visualization system designed to enhance the understanding of quantum encoders and their role in QNNs. 
To obtain the encoded data before the optimization process, we introduced \textit{Encoder Expectation Measurement}, which can extract and convert the intermediate encoded states into scalar values.
Furthermore, from the perspective of quantum, we presented the approach to intuitively analyze
whether the encoder effectively distinguishes the features of different classes. 
We also conducted case studies and expert interviews with domain users, whose feedback highlights the effectiveness and usability of \toolName\ in addressing the challenges of encoders' explainability.

Two key takeaways from the evaluation of \toolName{} are: First, designing encoders whose features closely align with the training dataset can reduce the optimization burden, effectively simplifying the learning process. 
This finding aligns with the insights collected from the analysis of the Encoder Map View.
Second, encoding data points in a way that ensures the distinguishability of data points with different classes, as visualized by the Quantum Distribution Map, leads to smoother convergence and better performance. 

\acknowledgments{%
	This project is supported by the Ministry of Education, Singapore, under Academic Research Fund Tier 2 (Proposal ID: T2EP202220049). Any opinions, findings and conclusions, or recommendations expressed in this material are those of the author(s) and do not reflect the views of the Ministry of Education, Singapore.
}

\bibliographystyle{abbrv-doi-hyperref}

\bibliography{template}

\newpage
\clearpage

\appendix % You can use the `hideappendix` class option to skip everything after \appendix

\onecolumn

\section{Interviews in the Formative Study}
\label{sec:appen4question}

To ensure the design of our visualization system effectively meets the needs of domain experts in quantum computing, we conducted a structured formative study aimed at uncovering key challenges and opportunities in analyzing quantum data encoders. This study focused on gathering insights into workflows, interpretability issues, and usability concerns associated with quantum data encoders.

The interviews provided a comprehensive understanding of how experts approach encoder design, evaluate encoded feature maps, and analyze the intermediate quantum states during the encoding process. These discussions highlighted the importance of bridging the gap between classical and quantum neural networks, supporting flexible experimentation with diverse datasets, and enabling intuitive visualizations for performance analysis and debugging.
Table \ref{table:3} presents the key questions designed to elicit feedback from participants, focusing on their expectations for visualization systems that enhance interpretability, usability, and adaptability in quantum data encoding workflows.

\begin{table}[htbp]
\caption{%
The pre-defined questions used in the preliminary study for the session of the design requirement collection. 
}
\centering
% \begin{tabular}{c|p{0.8\columnwidth}}
\begin{tabular}{c|p{0.85\columnwidth}}
\hline
Q1                      & What aspects of Quantum Neural Networks do you find most challenging to understand or implement?                          \\ \cline{2-2} 
Q2                      & How do you currently approach designing or selecting encoders for QNNs?                        \\ \cline{2-2} 
Q3                      & What types of visual aids would help you better analyze the performance of QNNs and encoders? \\ \cline{2-2} 
Q4                      & How do you interpret the outputs of QNNs, and what tools or methods do you use in this process?                                         \\ \cline{2-2} 
Q5                      & What features should a visualization system include to make encoder designs more accessible and intuitive?                                                \\ \cline{2-2} 
Q6                      & In what ways can visualization tools improve your ability to debug or optimize QNN architectures?                                                \\ \cline{2-2} 
Q7                      & What strategies or features can reduce the learning curve for domain users working with QNNs and encoders?     \\ \cline{2-2} 
\multicolumn{1}{l|}{Q8} & How can visualization systems enhance the interpretability of QNNs for both researchers and practitioners?                             \\ \hline
\end{tabular}
\label{table:3}
\end{table}

\section{System Interface}

At the forefront of the interface lies a prominent header, designed to capture the user's attention and provide context for the exploration of quantum neural networks and data encoding mechanisms. 
The central interface comprises several interconnected components that collectively illustrate the process of quantum data encoding and neural network training. These components are arranged to provide a logical flow of information, beginning with the presentation of classical data and culminating in the visualization of trained quantum neural network outputs. 
The interface integrates an interactive control panel at its uppermost section, enabling users to dynamically adjust critical parameters such as the encoding method, training epochs, learning rate, and the search space for optimization. 
To ensure users have a foundational understanding of the underlying principles, the system incorporates a comprehensive tutorial on quantum neural networks (QNNs) and quantum data encoders. This tutorial is designed to bridge the gap between classical machine learning concepts and quantum computing paradigms, enabling users to grasp the unique advantages and challenges associated with quantum neural networks.
A specific focus is placed on angle encoding, a widely used method for quantum data representation. This technique utilizes the rotation angles of quantum gates, such as $R_x$, $R_y$, or $R_z$, to encode features of classical data. For instance, a feature value $x_i$ can be represented as $R_y(2x_i)|0\rangle$, allowing for efficient integration into quantum circuits.

\begin{figure}[htbp]% specify a combination of t, b, p, or h for top, bottom, on its own page, or here
  \centering % avoid the use of \begin{center}...\end{center} and use \centering instead (more compact)
  \includegraphics[width=\linewidth]{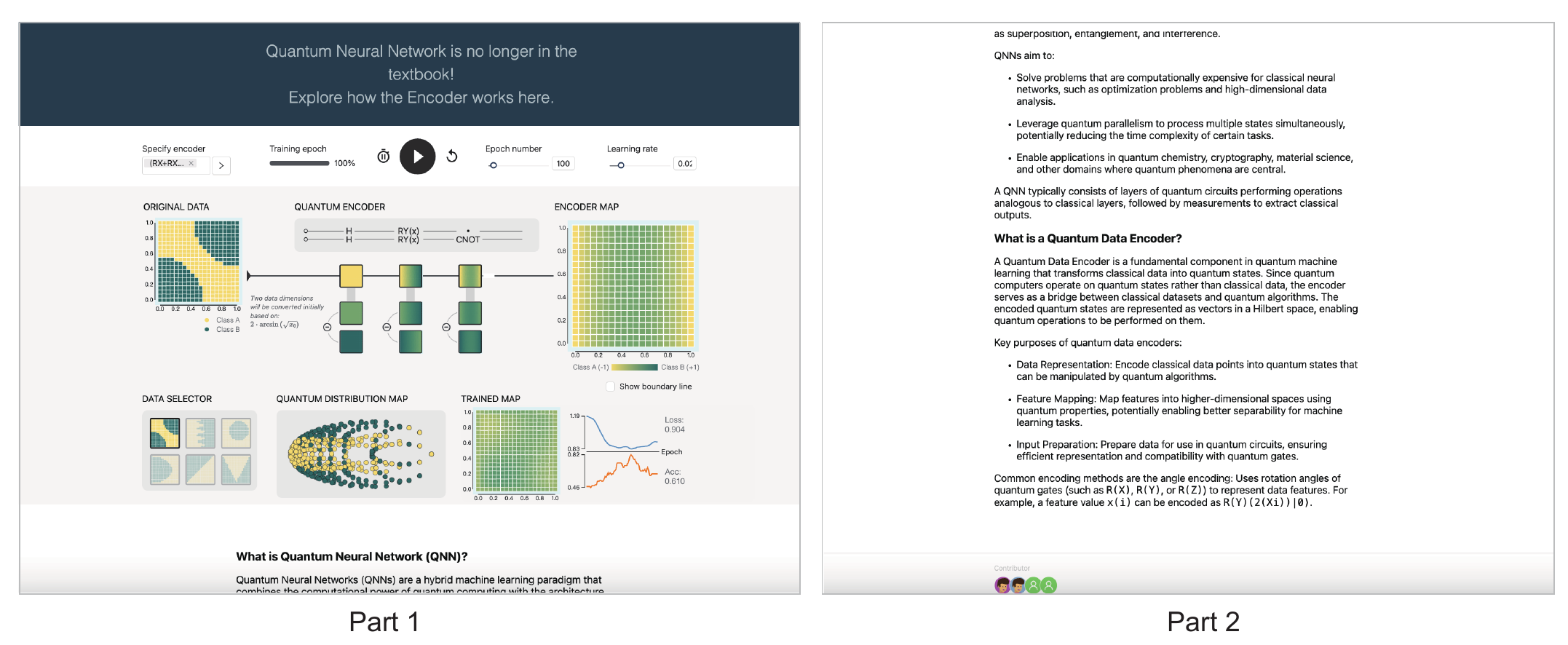}
  \caption{The screenshot of the visualization system, which consists of three components, i.e., the header, the system interface, and the tutorial.}
  \label{fig:appen1}
\end{figure}

\end{document}